\documentclass[aps,twocolumn,preprintnumbers,footinbib,prl,superscriptaddress,10pt,nofootinbib]{revtex4-1}

\makeatletter
\def\l@subsubsection#1#2{}
\def\l@subsubsubsection#1#2{}
\makeatother

\usepackage[utf8]{inputenc}

\usepackage{graphicx,amssymb,amsmath,amsthm,amsfonts,epsfig,epsf}
\usepackage[usenames,dvipsnames,svgnames,table]{xcolor}
\usepackage{epstopdf}
\definecolor{darkred}{rgb}{0.5,0,0}
\usepackage{aas_macros}
\usepackage{bm}
\usepackage{dcolumn}
\usepackage{latexsym}
\usepackage{rotating}
\usepackage{longtable}

\usepackage{enumerate}
\usepackage{tensor,multirow}
\usepackage{url}
\usepackage[linktocpage]{hyperref}

\def\be{\begin{equation}}
\def\ee{\end{equation}}
\newcommand{\beq}{\begin{eqnarray}}
\newcommand{\eeq}{\end{eqnarray}}

\def\ba{\begin{align}}
\def\ea{\end{align}}

\begin{document}

\title{Moving black holes:\\
energy extraction, absorption cross-section and the ring of fire}
\author{
Vitor Cardoso
}
\affiliation{Centro de Astrof\'{\i}sica e Gravita\c c\~ao - CENTRA, Departamento de F\'{\i}sica, Instituto
  Superior T\'ecnico - IST, Universidade de Lisboa - UL, Avenida Rovisco Pais 1, 1049-001 Lisboa, Portugal}
\affiliation{Theoretical Physics Department, CERN 1 Esplanade des Particules, Geneva 23, CH-1211, Switzerland}
\author{Rodrigo Vicente}
\affiliation{Centro de Astrof\'{\i}sica e Gravita\c c\~ao - CENTRA, Departamento de F\'{\i}sica, Instituto
	Superior T\'ecnico - IST, Universidade de Lisboa - UL, Avenida Rovisco Pais 1, 1049-001 Lisboa, Portugal}

\begin{abstract}
We consider the interaction between a plane wave and a (counter-moving) black hole. We show that
energy is transferred from the black hole to the wave, giving rise to a {\it negative} absorption cross-section. 
Moving black holes absorb radiation {\it and} deposit energy in external radiation.
Due to this effect, a black hole hole of mass $M$ moving at relativistic speeds in a cold medium will appear to be surrounded
by a bright ``ring'' of diameter $3\sqrt{3}GM/c^2$ and thickness $\sim GM/c^2$. 
\end{abstract}

\maketitle

%%%%%%%%%%%%%%%%%%%%%%%%%%%%%%%%%%%%%%%%%%%%%%%%%%%%
%\section{Introduction}\label{introduction}
%%%%%%%%%%%%%%%%%%%%%%%%%%%%%%%%%%%%%%%%%%%%%%%%%%%%
%%%%%%%%%%%%%%%%%%%%%%%%%%%%%%%%%%%%%%%%%%%%%%%%%%%%%%%%%%%%%%%%%%%%%%%%%%%%%
\noindent{\bf{\em I. Introduction.}}
%%%%%%%%%%%%%%%%%%%%%%%%%%%%%%%%%%%%%%%%%%%%%%%%%%%%%%%%%%%%%%%%%%%%%%%%%%%%%
The response of a black hole (BH) to an incoming wave has been studied for decades, in the frame where the BH is at rest~\cite{Matzner,Starobinski2:1973,Teukolsky:1974yv,Unruh:1976fm,Sanchez:1977si,MTB,Glampedakis:2001cx,Macedo:2013afa,Crispino:2009xt,Leite:2016hws,Leite:2017zyb,Leite:2018mon,Benone:2018rtj}.
Such interaction is crucial to understand how BHs react to their environment, what types of signatures are imprinted
by strong-field regions and their possible observational effects.
It was shown that non-spinning BHs absorb low-frequency plane waves. For a BH of mass $M$, the low-frequency absorption cross-section of scalars is equal to the horizon area, $\sigma=16\pi (GM/c^2)^2$. High frequency plane waves, on the other hand, are absorbed with a cross-section $\sigma=27\pi (GM/c^2)^2$~\cite{MTB,Macedo:2013afa,Das:1996we}. Although spinning BHs also absorb plane waves, they can amplify certain, low-frequency, angular modes through superradiance~\cite{zeldovich1,zeldovich2,Brito:2015oca} (which also acts on charged BHs~\cite{Benone:2019all}). Superradiance extracts energy away from such BH and provides important signatures of possible fundamental ultralight fields in nature~\cite{Brito:2015oca,Arvanitaki:2016qwi,Brito:2017wnc,Ikeda:2019fvj}.

A significant fraction of BHs are found in binaries, such as those seen by the LIGO/Virgo observatories~\cite{LIGOScientific:2018mvr}.
In addition, most BHs are moving at high speeds relative to our own frame. Thus, an understanding of the interaction between waves and moving BHs is a necessary ingredient to explore the enormous potential of such sources~\cite{Bernard:2019nkv,Wong:2019kru}. 

It was recently pointed out that BH binaries could amplify incoming radiation through a gravitational slingshot mechanism for light~\cite{Bernard:2019nkv}. The argument requires only {\it one} BH moving with velocity $v$, and a photon reflected at an angle of $180 ^{\circ}$ by the strong-field region (such orbits do exist~\cite{MTB}). Then, a trivial change of frames and consequent blueshift yields
\be
E_f^{\rm peak}=\frac{1+v/c}{1-v/c}\,E_i\,,\label{max_amp}
\ee
for the energy gain by the photon during the process. This is also the blueshift by photons reflecting off a mirror moving with velocity $v$. In addition, effective field theory methods were recently used to suggest that BH binaries could amplify radiation through superradiance~\cite{Wong:2019kru}. Again, the argument seems to imply that a single moving BH is able to amplify incoming radiation.

The purpose of the present work is to study the scattering of a plane wave off a moving BH. Clearly, such study involves ``only'' a Lorentz transformation of the well-known results for BHs at rest. Our purpose is to generalize to BH physics the classical problem of scattering off a moving mirror or a sphere, addressed by Sommerfeld and others~\cite{Sommerfeld:1964,Restrick:1968}.
Our results are surprisingly simple but non-trivial, interesting and -- as far as we are aware -- new. 

%%%%%%%%%%%%%%%%%%%%%%%%%%%%%%%%%%%%%%%%%%%%%%%%%%%%%%%%%%%%%%%%%%%%
%\section{Amplification in the weak field regime}
%%%%%%%%%%%%%%%%%%%%%%%%%%%%%%%%%%%%%%%%%%%%%%%%%%%%%%%%%%%%%%%%%%%%
%%%%%%%%%%%%%%%%%%%%%%%%%%%%%%%%%%%%%%%%%%%%%%%%%%%%%%%%%%%%%%%%%%%%%%%%%%%%%
\noindent{\bf{\em II. Amplification in the weak field regime.}}
%%%%%%%%%%%%%%%%%%%%%%%%%%%%%%%%%%%%%%%%%%%%%%%%%%%%%%%%%%%%%%%%%%%%%%%%%%%%%
Consider a BH of mass $M$ and a high-frequency photon, described by null geodesics in the BH spacetime, with a large impact parameter $b\gg M$ and moving in the $-z$ direction.
The photon energy is $E_i$ in the frame where the BH is moving in the $+z$ direction with velocity $v$.
A boost in the $+z$ direction brings us to the BH frame, and blueshifts the wave to $E_1=\sqrt{(1+v/c)/(1-v/c)}E_i$.
In this frame, the photon is deflected by the Einstein angle $\alpha=4GM/(bc^2)$. Now boost back to the $-z$ direction, where due to relativistic aberration the angle with the $z-$axis is $\alpha'\sim \alpha\sqrt{(1+v/c)/(1-v/c)}$, and the frequency is now
$E_f=E_1/(\gamma(1+v\cos\alpha'/c))$. One finds the weak field energy amplification for such photons
\be
E_f^{\rm weak}=\left(1+\frac{8G^2M^2v}{b^2c^4(c-v)}\right)E_i\,.\label{deflection_weak}
\ee

If a plane wave is passing through, one can see that the $1/r$ nature of the gravitational potential causes the total extracted energy to diverge; this phenomenon is akin to the divergence of the scattering cross section of the Coulomb potential~\cite{Merzbacher}. 
For a body of size $R_{\rm min}$ moving in a plane wave of density $\rho$ and extent $R_{\rm max}$, we find the total energy loss per second
\be
d E/dt=-\frac{16\pi G^2M^2v^2\rho}{c^2(c-v)}\log{\left(R_{\rm max}/R_{\rm min}\right)}\,.
\ee
%

%%%%%%%%%%%%%%%%%%%%%%%%%%%%%%%%%%%%%%%%%%%%%%%%%%%%%%%%%%%%%%%%%%%%%%%%%%%%%
\noindent{\bf{\em III. Amplification in the strong-field regime.}}
%%%%%%%%%%%%%%%%%%%%%%%%%%%%%%%%%%%%%%%%%%%%%%%%%%%%%%%%%%%%%%%%%%%%%%%%%%%%%
%
\begin{figure}
\begin{tabular}{c}
\includegraphics[width=7.5cm,height=7.5cm,keepaspectratio]{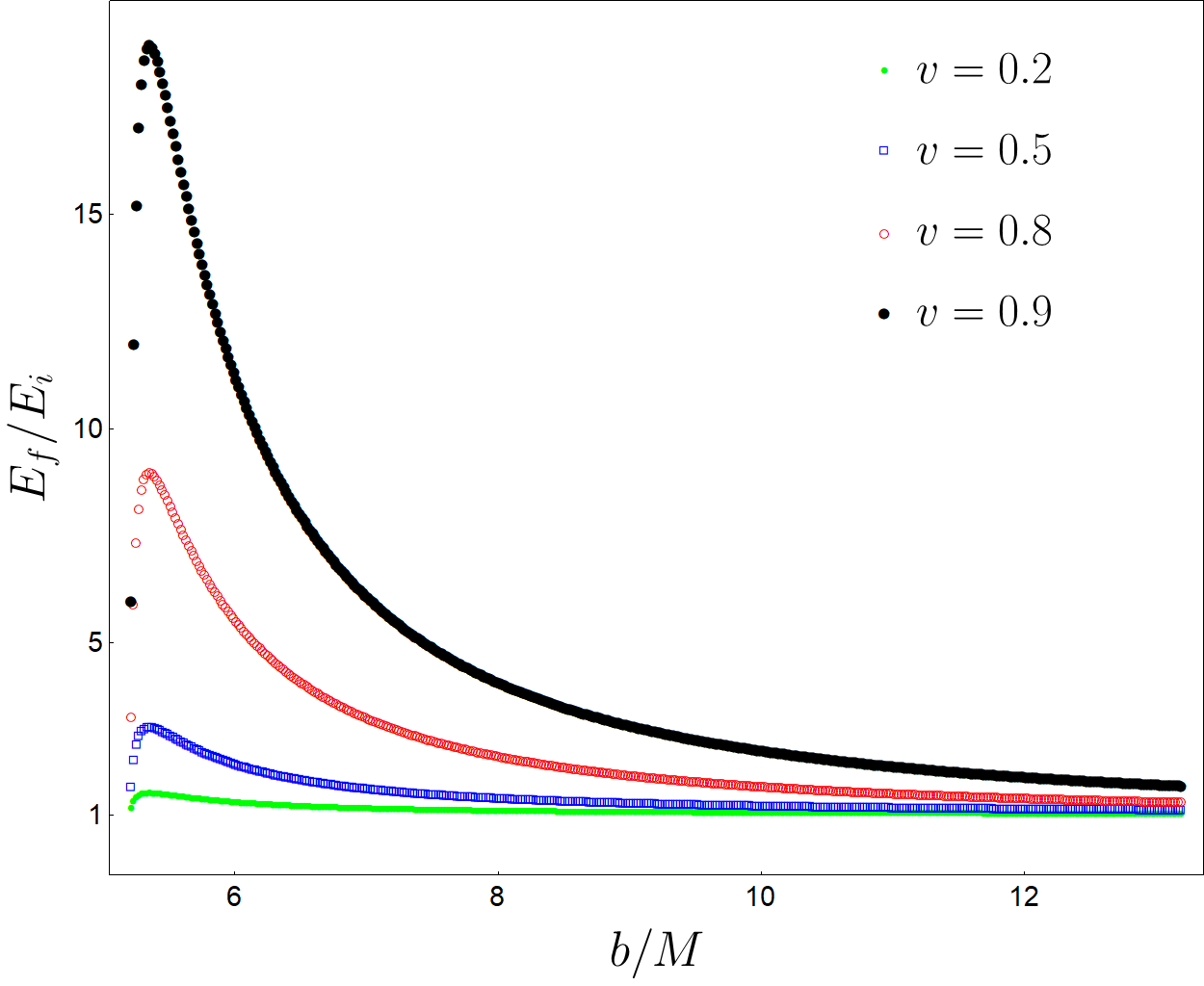} 
%&
%\includegraphics[width=8.3cm,height=6cm,keepaspectratio]{Plots/field_decay0p1.png} 
\end{tabular}
\caption{Energy gain of a (high frequency) photon scattered off a moving BH. The photon has initial energy $E_i$, impact parameter $b$ and scatters off a BH moving with velocity $v$ in the opposite direction. The final energy is $E_f$. The peak of each curves agrees, to numerical precision, with Eq.~\eqref{max_amp}. For impact parameter $b<3\sqrt{3}M$, the photon is absorbed by the BH.}
\label{fig:amplification_null}
\end{figure}
The energy {\it loss} of radiation when a body travels through a co-propagating stream is {\it smaller} than the energy {\it gain} when counter-propagating. This effect gets more pronounced at relativistic speeds and in the strong-field regime. We therefore focus exclusively on amplification by counter-moving waves.
The strong-field regime gives rise to large deflections in the photon's trajectory, and consequent large energy amplification, with a peak value
described by Eq.~\eqref{max_amp}. Henceforth, for simplicity, we use units with $c=G=1$. To compute rigorously the energy amplification at all impact parameters, we study null geodesics in the spacetime of a moving BH. In isotropic coordinates the Schwarzschild metric is given by
\be
ds^2=-\frac{(1-A)^2}{(1+A)^2}dt^2+(1+A)^4\left(dx^2+dy^2+dz^2\right)\,,
\ee
where $A=M/(2\rho)$ and $\rho^2=x^2+y^2+z^2$. Here, the standard Schwarzschild radial coordinate is related with $\rho$ via $r=\rho(1+A)^2$. Perform a boost along the $z$ direction, by letting 
\be
\hat{t}=\gamma (t+v z)\,,\quad \hat{z}=\gamma(z+vt)\,,\quad \hat{y}=y\,,\quad \hat{x}=x\,.\label{boost}
\ee
This yields the metric describing a BH moving with velocity $v$ and Lorentz factor $\gamma^2=1/(1-v^2)$.

It is now a simple question to study the scattering of a plane wave of null particles: follow initially counter-moving null geodesics, of impact parameter $b$ (i.e., null geodesics with $\hat{y}(\hat{t}=0)=b$ and $\dot{\hat{x}}=\dot{\hat{y}}=0$ at large distances) and monitor their energy $E=v^\mu\,p_\mu$, where $p$ is the four-momentum associated with the geodesic and $v^\mu=(1,0,0,0)$ the four-velocity of the observer.

Our results are shown in Fig.~\ref{fig:amplification_null} for different velocities $v$. There is a minimum impact parameter $b=3\sqrt{3}M$, below which the photon simply falls onto the BH. As we increase the impact parameter starting from this value, the energy gain peaks very rapidly at a value precisely (to within numerical precision) described by Eq.~\eqref{max_amp}: these are photons which are reflected back by the geometry. There are in fact a multitude of impact parameters for which photons are reflected back: for 
\beq
b/M&=&b_1/M=5.356\pm 0.003\,,\\
b/M&=&b_2/M=5.199\pm 0.002\,,
\eeq
the photon circles the BH exactly half an orbit (with a distance of minimum approach of $r/M=3.521\pm 0.001$) and one-and-a-half orbit (with a distance of minimum approach of $r/M=3.001\pm 0.001$), respectively; for impact parameters closer to the critical value a larger number of orbits around the BH are possible. At large impact parameters, our numerical results are well described by the weak-field result~\eqref{deflection_weak}.

%%%%%%%%%%%%%%%%%%%%%%%%%%%%%%%%%%%%%%%%%%%%%%%%%%%%%%%%%%%%%%%%%%%%
\noindent{\bf{\em IV. The high-frequency absorption cross-section.}}
%%%%%%%%%%%%%%%%%%%%%%%%%%%%%%%%%%%%%%%%%%%%%%%%%%%%%%%%%%%%%%%%%%%%
\begin{table}[b]
\begin{tabular}{cc||cc}
\hline
\hline
$v$ & $\sigma^{\rm abs}_{20}/(\pi M^2)$ &$v$ & $\sigma^{\rm abs}_{20}/(\pi M^2)$    \\ 
\hline
\hline
0.00 & 27.0                             &0.30  & 2.1  \\
0.01 & 26.4                             &0.50  &-31.1\\
0.02 & 25.8                             &0.80  &-205.6\\
0.10  & 20.6                            &0.90  &-496.8\\
 \hline
\hline
\end{tabular}. 
\caption{Absorption cross-section for a BH moving with velocity $v$ onto a constant flux wave. The incoming wave has a finite spatial extent in the direction transversal to the motion, forming a cylinder of radius $R=20M$. Notice that the absorption cross-section becomes negative at large velocities, indicating that BH transfers energy to the scattered waves.}
\label{table_basis}
\end{table}
In a scattering experiment, where a plane wave hits a moving BH head-on, one can define an absorption cross-section
\be
\sigma^{\rm abs}=\frac{E_{\rm in}-E_{\rm out}}{E_{\rm in}/A_{\rm in}}\,,
\ee
where $E_{\rm in}$ is the total energy in the plane wave, $E_{\rm out}$ is the total energy in the outgoing wave after interaction with the BH, and $A_{\rm in}$ is the surface area that the incident plane occupies. As we showed, due to the long-range character of gravity, the absorption cross-section above diverges~\cite{Merzbacher}. We define instead
a finite quantity $\sigma^{\rm abs}_{20}$, computed by sending a constant flux wave centered at the BH, but with finite transverse size of radius $R=20M$. This quantity is shown in Table~\ref{table_basis} for null particles.

The cross-section $\sigma^{\rm abs}_{20}$ is to a good approximation equal to its geometric-optics counterpart $\sigma=27\pi M^2$
for BHs at rest. It starts decreasing when the BH moves for the reasons discussed previously: photons with impact parameter $b=b_1$ are given energy. Our results seem to be well described by 
\be
\frac{\sigma^{\rm abs}_{20}}{\pi\,M^2}\sim 27-a_1v-a_2\frac{v(1+v)}{(1-v)}\,,\label{fit_cross}
\ee
with $a_1=28.8,\, a_2=29.1$, which reproduces the numerical points between $v=[0,0.99]$ to within $1\%$ accuracy. 
The coefficients $a_1,\, a_2$ grow when the incident surface radius $R$ grows (for example, a calculation of $\sigma^{\rm abs}_{10}$, for $R=10$, leads to $a_1=18.0,\,a_2=20.0$; such cross section also becomes negative at large $v$). 

From the previous discussion, it could also be anticipated that the absorption cross section becomes negative at large enough velocities.
Given that the only scale in the traverse directions is that of the BH, $M$, photons with an impact parameter $b\sim b_1$ within a width $\sim M$ will also be amplified. On the other other hand, all photons with impact parameter smaller than $3\sqrt{3}M$ are absorbed. Thus, the cross absorption section is expected to be of order $\sim 27 \pi M^2-2\pi \times 3\sqrt{3}M^2\left((1+v)/(1-v)-1\right)$. For large velocities, this (order-of-magnitude) argument predicts a negative cross section $\sim -10\pi M^2 (1+v)/(1-v)$, in rough agreement with the numerical fit~\eqref{fit_cross}.

%%%%%%%%%%%%%%%%%%%%%%%%%%%%%%%%%%%%%%%%%%%%%%%%%%%%%%%%%%%%%%%%%%%%%%%%%%%%%
\noindent{\bf{\em V. The absorption cross-section of moving BHs.}}
%%%%%%%%%%%%%%%%%%%%%%%%%%%%%%%%%%%%%%%%%%%%%%%%%%%%%%%%%%%%%%%%%%%%%%%%%%%%%
%
\begin{figure}
\begin{tabular}{c}
\includegraphics[width=7.5cm,height=7.5cm,keepaspectratio]{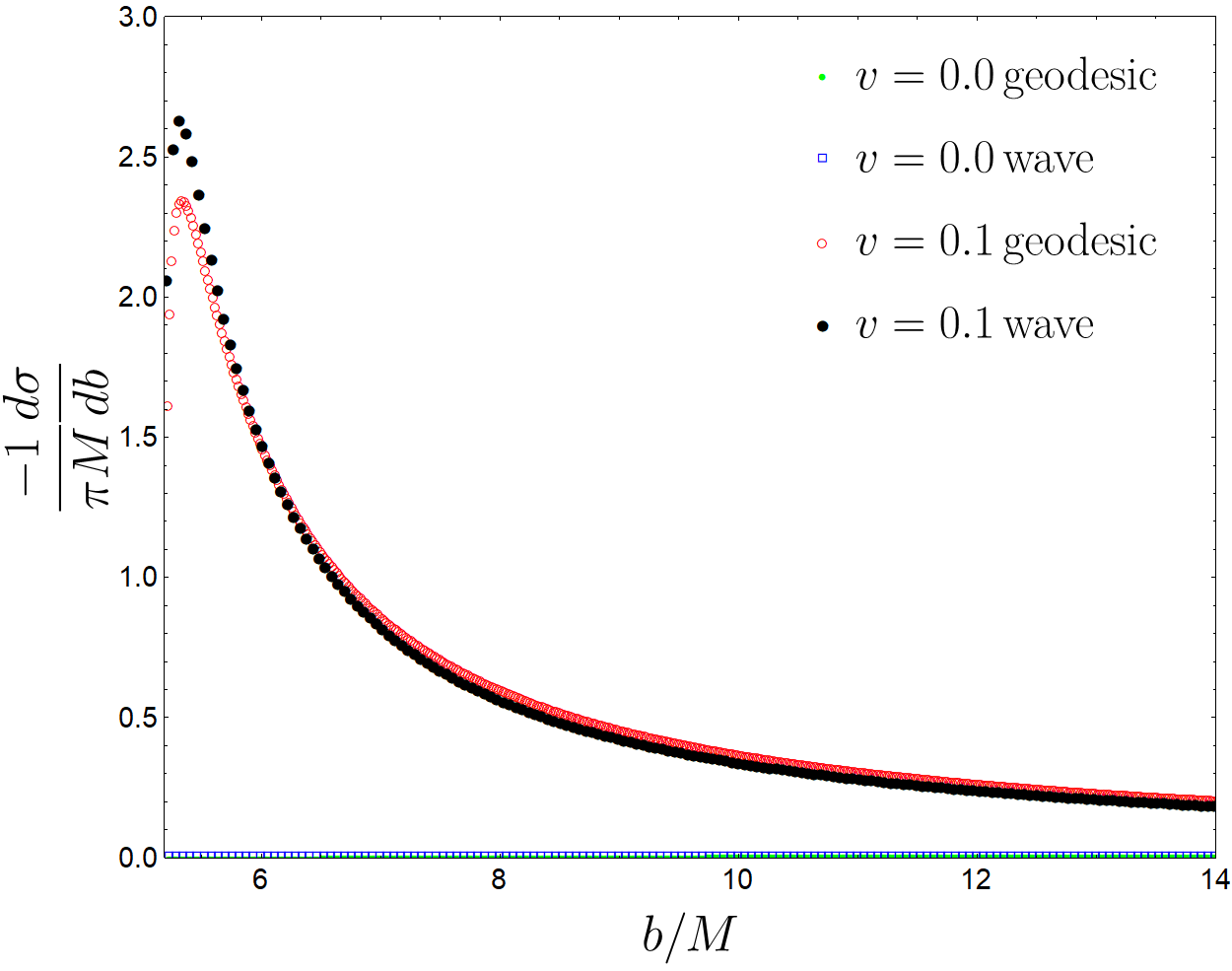} 
%&
%\includegraphics[width=8.3cm,height=6cm,keepaspectratio]{Plots/field_decay0p1.png} 
\end{tabular}
\caption{Numerical values of the quantity $\frac{d \sigma}{d b}$ obtained through (null geodesic) geometric-optics, and wave-scattering with $\omega M =17$. The two methods are in clear agreement for $b\gg M$.}
\label{fig:dsigma}
\end{figure}
Consider now the extension of the previous results to arbitrary low frequency waves, where geometric optics no longer provides an adequate description of the scattering phenomenon. Let us focus on a minimally coupled scalar field theory described by the action
\be
S= \int d^4x \sqrt{-g} \,g_{\mu \nu} \nabla^\nu\phi(\nabla^\mu\phi)^* \,.
\ee
%
%Consider first the 
%\begin{equation} \label{scalarfec}
%\Box \phi=0 \,.
%\end{equation}
%
For most situations of interest, the scalar field is but a small perturbation and can be studied in a fixed spacetime geometry -- the so-called test field approximation.
A well-studied problem concerns the scattering on a Schwarzschild geometry describing a BH at rest. In this setup, consider the ansatz
\begin{equation} \label{ansatzscalar}
\phi_{\omega_1} (t, r, \theta, \varphi)=\sum_{l, m}K_l^m e^{-i \omega_1 t}\, Y_l^{m}(\theta, \varphi) \frac{\psi_{\omega_1}(r)}{r} \,,
\end{equation} 
where $K_l^m$ are complex coefficients, $Y_l^m$ the spherical harmonics, and the radial function $\psi_{\omega_1}$ also carries an $(l,m)$ dependence. 
Define $f\equiv 1-2M/r$ and a tortoise coordinate $r_*$ satisfying $dr/dr_*=f$. The massless Klein-Gordon equation yields 
%%
%\beq 
%\label{scalartifec}
%&&f^2 \frac{d^2}{d r^2} \psi_{\omega_1} + f f'\frac{d}{d r} \psi_{\omega_1}+\left(\omega_1^2 -V\right) \psi_{\omega_1}=0\,,\\
%%
%&&V=f\left(\frac{l(l+1)}{r^2}+\frac{2M}{r^3} \right) \,,
%\eeq
%%
%
\be
\frac{d^2}{d r_*^2} \psi_{\omega_1} +\left[\omega_1^2 -f\left(\frac{l(l+1)}{r^2}+\frac{2M}{r^3} \right)\right] \psi_{\omega_1}=0\,.\label{scalartifec}
%
%&&V=f\left(\frac{l(l+1)}{r^2}+\frac{2M}{r^3} \right) \,,
\ee
This equation admits the asymptotic solution 
\begin{equation} \label{incscalar}
\psi_{\omega_1}(r \to +\infty)\sim I \, e^{-i \omega_1 r_*}+ R \, e^{+i \omega_1 r_*} \,,
\end{equation} 
with $I$ and $R$ the complex-valued amplitudes of the incident and reflected spherical waves, respectively.
The transmitted spherical wave at the BH horizon can be written as
\be
\psi_{\omega_1}(r \to 2M)\sim T\, e^{-i \omega_1 r_*}\,.\label{transscalar}
\ee
%
%where the sign of the exponential argument was chosen in such a way that the wave $e^{-i\omega_1 (t + r_*)}$ has negative group velocity along $r$
%
The quantities $(R/I)$ and $(T/I)$ appearing in the above solution are complex functions of $l$ and $\omega_1 M$, but, for simplicity, we omit this in our notation. 

%The solutions of the form of Eq.~\eqref{ansatzscalar} with asymptotic behaviour~\eqref{incscalar} and~\eqref{transscalar} describe the %scattering of scalar waves by a static BH at rest.
It is well-known that a (distorted) plane wave can be written as a partial wave expansion, which asymptotically reads~\cite{Matzner,landau1981quantum}
\begin{align}
e^{-i \omega_1(t+ z_*)}\simeq  -\sum_{l, m} \frac{e^{-i \omega_1 (t+r_*)}}{2 i \omega_1  r} &\sqrt{4\pi (2l+1)} Y_l^0(\theta,\varphi)\\
&+ \text{outgoing wave} \nonumber \,,
\end{align}
where $z_* \equiv r_* \cos \theta$.\\
Choosing $K_l^m\equiv - \delta_0^m\sqrt{4\pi (2l+1)}/(2 i \omega_1 I)$, one can rewrite the asymptotic behaviour of $\phi_{\omega_1}$ as
\begin{equation}
\phi_{\omega_1}(r\to+\infty)\sim e^{-i \omega_1 (t+ z_*)}+\sum_{l, m} \frac{\mathcal{R}}{r} Y_l^m e^{-i \omega_1 (t -  r_*)}\,,\label{incscalarplane}
\end{equation}
with $\mathcal{R}$ a complex function of $R/I$. Thus, with this choice of $K_l^m$, Eq.~\eqref{ansatzscalar} describes the scattering of scalar plane waves (propagating along the $-z$ direction) by a BH at rest.

%Noether's particle-number current is given by
%%
%\begin{equation}
%j^\mu=- \frac{i}{2}\,[ \phi^* \nabla^\mu \phi - \phi\, (\nabla^\mu \phi)^*] \,.
%\end{equation}
%%
%Thus, the net number of particles $\mathcal{N}$ (measured by a stationary observer at infinity) crossing a spherical surface $S_r$ of radius $r$ per unit time is
%%
%\begin{equation} \label{scalarfluxc}
%\partial_{t}\mathcal{N} =-\int_{S_{r}} d\Omega\, r^2 j^{r} \,.
%\end{equation} 
%%
%In the limit $r\to +\infty$, this gives
%%
%\begin{equation}
%\partial_{t}\mathcal{N}= \sum_{l}\frac{\pi(2l+1)}{\omega_1} \left(1-\left|\frac{R_l(\omega_1)}{I_l(\omega_1)}\right|^2\right)\,.
%\end{equation}
%%
The stress-energy tensor of a massless scalar field is
\begin{equation}
T_{\mu \nu}= \frac{1}{2}\,\left( \partial_\mu \phi^* \partial_\nu \phi + \partial_\mu \phi\, \partial_\nu \phi^*\right)- \frac{1}{2}\,g_{\mu \nu} \partial^\alpha \phi^* \partial_\alpha \phi \,.
\end{equation}
So, the net energy $\mathcal{E}$ (measured by a stationary observer at infinity) entering a spherical surface $S_r$ of radius $r$, per unit of time, is
\begin{equation} \label{scalarfluxc}
\partial_{t}\mathcal{E} =\int_{S_{r}} d\Omega\, r^2 T_{t r} \,.
\end{equation} 
In the limit $r\to +\infty$, this gives
\begin{equation}
\partial_{t}\mathcal{E}= \sum_{l}\pi(2l+1) \left(1-\left|\frac{R_l(\omega_1 M)}{I_l(\omega_1 M)}\right|^2\right)\,.
\end{equation}
One can define the (energy) absorption cross section
\begin{equation}
\sigma^{\text{abs}} \equiv	\frac{\partial_{t}\mathcal{E}}{(\omega_1)^2}\,,
\end{equation}
where we have used that the energy density current of the incident plane waves is $(\omega_1)^2$. Numerical evaluation of the last expression shows that, for a BH at rest, the absorption cross section is $\sigma^\text{abs} \simeq 27 \pi M^2$, a well-known result~\cite{MTB,Das:1996we}.    

Focus now on the problem of a scalar plane wave scattering off a Schwarzschild BH moving with velocity $v$ along the $+z$ direction. For simplicity, let us consider that the wave is propagating along the $-z$ direction. The Lorentz transformation of a plane wave is a (Doppler shifted) plane wave; by the principle of covariance, applying a (global) Lorentz boost of velocity $v$ along the $-z$ direction to the solution~\eqref{ansatzscalar}, which describes the scattering of a plane wave of frequency $\omega_1$ by a BH at rest, one gets a solution of the equations of motion describing the scattering of a plane wave of frequency $\omega=\omega_1 \sqrt{1-v}/\sqrt{1+v}$ by a BH moving with velocity $v$. In mathematical terms: by applying the Lorentz boost \eqref{boost}
%\begin{align}
%	&t=\gamma(\hat{t}-v \hat{z})\,, \nonumber\\
%	&z=\gamma(\hat{z}-v \hat{t})\,, \nonumber\\
%	&(x,y)=(\hat{x},\hat{y})
%\end{align}
to Eq.~\eqref{ansatzscalar}, we get a solution of the Klein-Gordon equation, which, in spherical coordinates centered at the BH, 
$\hat{x}=\hat{r} \sin \hat{\theta} \cos \hat{\varphi}$, $\hat{y}=\hat{r} \sin \hat{\theta} \sin \hat{\varphi}$, $\hat{z}-v \hat{t}= \hat{r} \cos \hat{\theta}$,
has the asymptotic behaviour
\begin{equation}\label{incscalarplaneboost}
\phi_{\omega_1}(\hat{r}\to+\infty)\sim e^{-i \omega (\hat{t}+ \hat{z}_*)}+\sum_{l, m} \frac{\mathcal{R}}{r} Y_l^m e^{-i \omega_1 (t -  r_*)}\,,
\end{equation}
with
\beq
&&t=\frac{\hat{t}}{\gamma}- \gamma v \cos \hat{\theta}\, \hat{r}\,,\qquad (r, r_*)=\sqrt{\xi}(\hat{r},\hat{r}_*) \,, \nonumber\\
&&\cos \theta=\frac{\gamma \cos \hat{\theta}}{\sqrt{\xi}}\,, \,\, \quad \qquad \hat{z}_* \equiv v \hat{t}+ \hat{r}_* \cos \hat{\theta} \,,
\eeq
where
\beq
\hat{r}_*&\equiv& \hat{r}+\frac{2M}{\sqrt{\xi}(1-v)}\log\left(\frac{\hat{r} \sqrt{\xi}}{2 M}-1\right)\,, \nonumber \\
\xi &\equiv& 1+\left(\gamma^2-1\right) \cos^2\hat{\theta}\,. 
\eeq
Thus, this solution describes the scattering of a (distorted) plane wave of frequency $\omega$ by a moving BH. 

The asymptotic solution~\eqref{incscalarplaneboost} can also be written as
\begin{align}
\phi_{\omega_1} \sim &\sum_{l} \frac{i}{2 \omega_1} \frac{2 l +1}{\hat{r} \sqrt{\xi}} e^{-i \omega_1 \left(\frac{\hat{t}}{\gamma}-\gamma v \cos \hat{\theta}\,\hat{r}\right)} \times \nonumber \\ 
&P_l\left(\frac{\gamma \cos \hat{\theta}}{\sqrt{\xi}}\right)\left(e^{-i \omega_1 \hat{r}_* \sqrt{\xi}}+ \frac{R_l(\omega_1 M)}{I_l(\omega_1 M)}e^{i \omega_1 \hat{r}_* \sqrt{\xi}}\right)	\,,
\end{align}
with $P_l$ the Legendre polynomial of the first kind. After a laborious calculation, and using Ref.~\cite{tableofintegrals} for angular integrations, one finds
\beq 
\sigma^\text{abs}&=&\sum_{l}\frac{\pi (2 l+1)^2}{ \omega^2 \gamma v} P_l\left(\frac{1}{v}\right) Q_l\left(\frac{1}{v}\right) \left(1-\left|\frac{R_l(\omega_1 M)}{I_l(\omega_1 M)}\right|^2\right)\nonumber \\
&+&\sum_{l'<l} \frac{2 \pi }{\omega^2 \gamma v}(2 l'+1)(2 l+1)P_{l'}\left(\frac{1}{v}\right) Q_l\left(\frac{1}{v}\right)\times \nonumber \\
&&\left((-1)^{1+l'+l}+ \text{Re}\left[\frac{R_{l'}^*(\omega_1 M)}{I_{l'}^*(\omega_1 M)}\frac{R_{l}(\omega_1 M)}{I_l(\omega_1 M)}\right]\right) \,,\label{cross_section}
\eeq
where $Q_l$ is the Legendre polynomial of the second kind, and we have used that the particle-number current of the incident plane waves is $\omega$. There is angular mode mixing in $l$ due to the fact that the boosted spherical harmonics lose their orthogonality properties. Note that
\begin{equation*}
\lim_{v\to 0} \frac{1}{v}P_{l'}\left(\frac{1}{v}\right) Q_l\left(\frac{1}{v}\right)= \frac{\delta_{l'}^{l}}{2l+1} \,,
\end{equation*}
which shows that one recovers known results when $v\to 0$.

Numerical evaluation of Eq.~\eqref{cross_section} shows qualitatively the same behaviour obtained with geometric-optics. In particular, our results indicate that the second sum of Eq.~\eqref{cross_section} diverges logarithmically with $l$. Using the interpretation $b\simeq l/\omega_1$, which is valid for large $l \gg 1$, this can be restated as a divergence in the impact parameter $b$; a well-known consequence of the long range character of gravity. For a quantitative comparison, we consider high-frequency ($\omega M \gg 1$) plane waves, and truncate the sums in~\eqref{cross_section} at $l=\omega_1 R$; describing an incident beam with maximum impact parameter $R$ (see section {\em IV}). 
%With $M\omega =17$ and $R=20M$, our results agree qualitatively with the ones in Table~\ref{table_basis} (both in sign and order of magnitude), but not quantitatively. Inspired by Bohr's %correspondence principle, we think that this numerical difference is due to the contribution of small $l$'s (small impact parameters $b$), since the wave results should reproduce the particle %(geodesic) ones only in the limit $\omega M \gg 1$ and $l\gg1$. To test this hypothesis, 
We computed the (finite) quantity $d \sigma^{\rm abs}/d b$ for $b\gg M$, which for waves is approximated by
\begin{equation}
\frac{d \sigma^{\rm abs}}{d b}(l \omega_1) \sim \omega_1\left[ \sigma^\text{abs}(l+1)-\sigma^\text{abs}(l)\right] \,.
\end{equation}
As shown in Fig.~\ref{fig:dsigma}, the numerical values obtained for this quantity by the two approaches are in very good agreement for $b\gg M$, as one expects.

%%%%%%%%%%%%%%%%%%%%%%%%%%%%%%%%%%%%%%%%%%%%%%%%%%%%%%%%%%%%%%%%%%%%%%%%%%%%%
\noindent{\bf{\em VI. Appearance of a moving black hole.}}
%%%%%%%%%%%%%%%%%%%%%%%%%%%%%%%%%%%%%%%%%%%%%%%%%%%%%%%%%%%%%%%%%%%%%%%%%%%%%
%
\begin{figure}
\begin{tabular}{c}
\includegraphics[width=7.5cm,height=7.5cm,keepaspectratio]{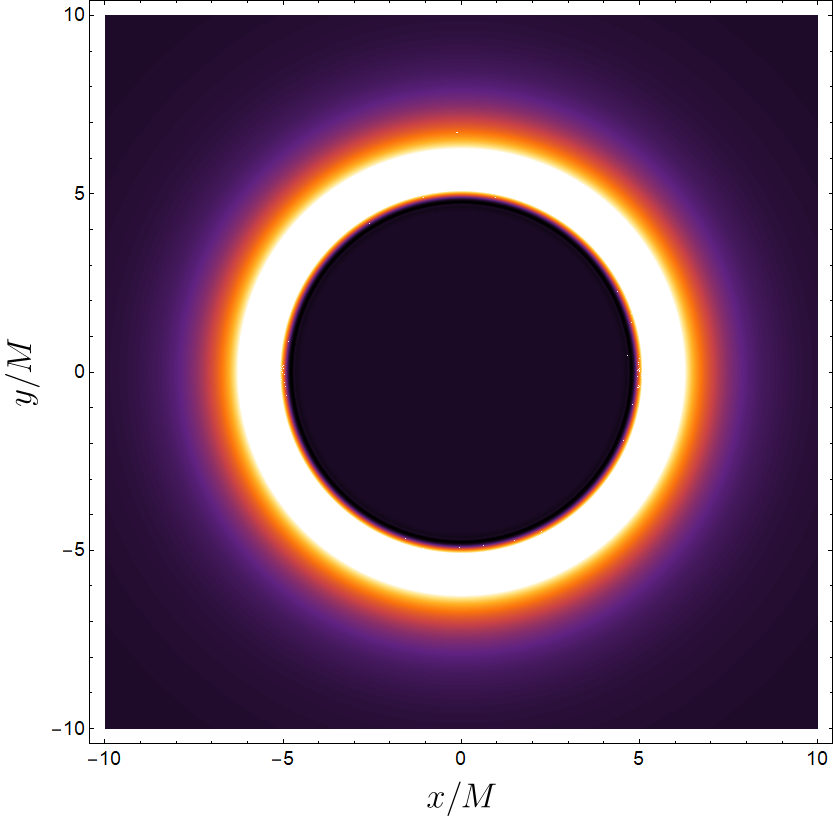} 
%&
%\includegraphics[width=8.3cm,height=6cm,keepaspectratio]{Plots/field_decay0p1.png} 
\end{tabular}
\caption{Appearance of a BH moving in a bath of cold (and counter-moving) radiation. The BH is moving along the $z$-axis towards us at a speed $v=0.9$. The colors denote energy flux intensity on a screen placed a short distance away from the BH. The peak energy flux is ten times larger than that of the environment. The bright ring has width $\sim M$ for all boost velocities $v$. For very large $v$ even a randomly-moving gas of photons will leave a similar observational imprint, since counter-moving photons will be red-shifted away.}
\label{fig:image}
\end{figure}
The large amplification for strongly-deflected photons implies that a rapidly moving BH looks peculiar. Downstream photons are deflected and blueshifted upstream. Thus a rapidly moving BH in a cold gas of radiation will be surrounded by a bright ring of thickness $\sim M$. A possible image of a moving BH is shown in Fig.~\ref{fig:image}. For a stellar-mass BH moving at velocities $v\sim 0.9996$ through the universe, the ambient microwave cosmic background will produce a kilometer-sized ring (locally $\sim 5000$ times hotter and brighter than the CMB) in the visible spectrum. 

%%%%%%%%%%%%%%%%%%%%%%%%%%%%%%%%%%%%%%%%%%%%%%%%%%%%%%%%%%%%%%%%%%%%
\noindent{\bf{\em Discussion.}}
%%%%%%%%%%%%%%%%%%%%%%%%%%%%%%%%%%%%%%%%%%%%%%%%%%%%%%%%%%%%%%%%%%%%
%\section{Discussion}
%%%%%%%%%%%%%%%%%%%%%%%%%%%%%%%%%%%%%%%%%%%%%%%%%%%%%%%%%%%%%%%%%%%%
The scattering of massless waves is a fundamental process in physics. 
We showed that the universal nature of gravity, together with the $1/r$ behavior of Newton's law
causes moving BHs to amplify plane waves, with a divergent cross-section. This is the only known example
of a negative absorption cross section of neutral fields.
We also showed that even a narrow beam of light can extract energy from a rapidly (counter-) moving BH.
These results apply to any massless wave in the high frequency regime.  
For BHs at rest, the absorption cross-section of low-frequency electromagnetic or gravitational waves vanishes, which may imply that
amplification happens sooner at low frequencies, for higher spins. This remains to be understood.
These results may have little practical application, since BHs are not expected to be traveling through our universe
at relativistic speeds: mergers of BHs or neutron stars lead at best to ``kicks'' in the remnant of $v\lesssim 10^{-2}$~\cite{Campanelli:2007ew,Brugmann:2007zj,Gonzalez:2007hi,Campanelli:2007cga} for astrophysical setups (even the high-energy merger of two BHs leads ``only'' to kicks of $v\lesssim 0.05$~\cite{Sperhake:2010uv}). For these velocities, the effects dealt with here
are only important when the BH moves in very extended media. Nevertheless, our results show how nontrivial strong gravity effects can be. 

On the other hand, the mechanism for energy extraction could be relevant in the context of fundamental light fields, with confined low-energy excitations~\cite{Bernard:2019nkv}. A BH binary in this setup could slow down and transfer some of its energy to the fundamental field, giving rise to potentially observable effects. However, since energy is being transferred to radiation of higher frequency, the process stalls eventually, since sufficiently energetic excitations are unbound.

The overall result of energy transfer to external radiation echoes that of the inverse Compton scattering for fast-moving electrons in a radiation field~\cite{Dolan:private,Beckmann}. In this latter process, a nearly-isotropic radiation field is seen as extremely anisotropic to the individual ultrarelativistic electrons. Relativistic aberration causes the ambient photons to approach nearly head-on; Thomson scattering of this highly anisotropic radiation reduces the electron's kinetic energy and converts it into inverse-Compton radiation by upscattering radio photons into optical or $X$-ray photons. The process we discuss here, involving BHs, is special: BHs are natural absorbers, but the universal -- and strong, close to the horizon -- pull of gravity can turn them also into overall amplifiers.

%%%%%%%%%%%%%%%%%%%%%%%%%%%%%%%%%%%%%%%%%%%%%%%%%%%%%%%%%%%%%%%%%%%%%%%%%%%%%
\noindent{\bf{\em Acknowledgments.}}
%%%%%%%%%%%%%%%%%%%%%%%%%%%%%%%%%%%%%%%%%%%%%%%%%%%%%%%%%%%%%%%%%%%%%%%%%%%%%
%
We are indebted to Sam Dolan and Jos\'e Nat\'ario for useful feedback and comments.
V.C. acknowledges financial support provided under the European Union's H2020 ERC 
Consolidator Grant ``Matter and strong-field gravity: New frontiers in Einstein's 
theory'' grant agreement no. MaGRaTh--646597. 
R.V.\ was supported by the FCT PhD scholarship SFRH/BD/128834/2017.
This project has received funding from the European Union's Horizon 2020 research and innovation programme under the Marie Sklodowska-Curie grant agreement No 690904.
We acknowledge financial support provided by FCT/Portugal through grant PTDC/MAT-APL/30043/2017.
The authors would like to acknowledge networking support by the GWverse COST Action 
CA16104, ``Black holes, gravitational waves and fundamental physics.''
%
%
% \end{acknowledgments}
%%%%%%%%%%%%%%%%%%%%%%%%%%%%%%%%%%%%%%%%%%%%%%%%%%%%%%%%%%%%%%%%%%%%%%%%%%%%%

%\newpage
\bibliographystyle{apsrev4}
\bibliography{References}

\begin{thebibliography}{36}%
\makeatletter
\providecommand \@ifxundefined [1]{%
 \@ifx{#1\undefined}
}%
\providecommand \@ifnum [1]{%
 \ifnum #1\expandafter \@firstoftwo
 \else \expandafter \@secondoftwo
 \fi
}%
\providecommand \@ifx [1]{%
 \ifx #1\expandafter \@firstoftwo
 \else \expandafter \@secondoftwo
 \fi
}%
\providecommand \natexlab [1]{#1}%
\providecommand \enquote  [1]{``#1''}%
\providecommand \bibnamefont  [1]{#1}%
\providecommand \bibfnamefont [1]{#1}%
\providecommand \citenamefont [1]{#1}%
\providecommand \href@noop [0]{\@secondoftwo}%
\providecommand \href [0]{\begingroup \@sanitize@url \@href}%
\providecommand \@href[1]{\@@startlink{#1}\@@href}%
\providecommand \@@href[1]{\endgroup#1\@@endlink}%
\providecommand \@sanitize@url [0]{\catcode `\\12\catcode `\$12\catcode
  `\&12\catcode `\#12\catcode `\^12\catcode `\_12\catcode `\%12\relax}%
\providecommand \@@startlink[1]{}%
\providecommand \@@endlink[0]{}%
\providecommand \url  [0]{\begingroup\@sanitize@url \@url }%
\providecommand \@url [1]{\endgroup\@href {#1}{\urlprefix }}%
\providecommand \urlprefix  [0]{URL }%
\providecommand \Eprint [0]{\href }%
\providecommand \doibase [0]{http://dx.doi.org/}%
\providecommand \selectlanguage [0]{\@gobble}%
\providecommand \bibinfo  [0]{\@secondoftwo}%
\providecommand \bibfield  [0]{\@secondoftwo}%
\providecommand \translation [1]{[#1]}%
\providecommand \BibitemOpen [0]{}%
\providecommand \bibitemStop [0]{}%
\providecommand \bibitemNoStop [0]{.\EOS\space}%
\providecommand \EOS [0]{\spacefactor3000\relax}%
\providecommand \BibitemShut  [1]{\csname bibitem#1\endcsname}%
\let\auto@bib@innerbib\@empty
%</preamble>
\bibitem [{\citenamefont {Matzner}(1968)}]{Matzner}%
  \BibitemOpen
  \bibfield  {author} {\bibinfo {author} {\bibfnamefont {R.A.} \bibnamefont
  {Matzner}}, }\href {\doibase 10.1063/1.1664470} {\bibfield  {journal}
  {\bibinfo  {journal} {\emph {Journal of Mathematical Physics}} }\textbf
  {\bibinfo {volume} {9}}, \bibinfo {pages} {163} (\bibinfo {year} {1968})},
  \Eprint {http://arxiv.org/abs/https://doi.org/10.1063/1.1664470}
  {https://doi.org/10.1063/1.1664470}\BibitemShut {NoStop}%
\bibitem [{\citenamefont {Starobinski} and \citenamefont
  {Churilov}(1973)}]{Starobinski2:1973}%
  \BibitemOpen
  \bibfield  {author} {\bibinfo {author} {\bibfnamefont {A.A.} \bibnamefont
  {Starobinski}} and \bibinfo {author} {\bibfnamefont {S.M.} \bibnamefont
  {Churilov}}, }\href@noop {} {\bibfield  {journal} {\bibinfo  {journal} {\emph
  {Zh. Eksp. Teor. Fiz.}} }\textbf {\bibinfo {volume} {65}}, \bibinfo {pages}
  {3} (\bibinfo {year} {1973})}, \bibinfo {note} {(Sov. Phys. - JETP, 38, 1,
  1973)}\BibitemShut {NoStop}%
\bibitem [{\citenamefont {Teukolsky} and \citenamefont
  {Press}(1974)}]{Teukolsky:1974yv}%
  \BibitemOpen
  \bibfield  {author} {\bibinfo {author} {\bibfnamefont {S.A.} \bibnamefont
  {Teukolsky}} and \bibinfo {author} {\bibfnamefont {W.H.} \bibnamefont
  {Press}}, }\href {\doibase 10.1086/153180} {\bibfield  {journal} {\bibinfo
  {journal} {\emph {Astrophys. J.}} }\textbf {\bibinfo {volume} {193}},
  \bibinfo {pages} {443} (\bibinfo {year} {1974})}\BibitemShut {NoStop}%
%%CITATION = ASJOA,193,443;%%
\bibitem [{\citenamefont {Unruh}(1976)}]{Unruh:1976fm}%
  \BibitemOpen
  \bibfield  {author} {\bibinfo {author} {\bibfnamefont {W.G.} \bibnamefont
  {Unruh}}, }\href {\doibase 10.1103/PhysRevD.14.3251} {\bibfield  {journal}
  {\bibinfo  {journal} {\emph {Phys. Rev.}} }\textbf {\bibinfo {volume} {D14}},
  \bibinfo {pages} {3251} (\bibinfo {year} {1976})}\BibitemShut {NoStop}%
%%CITATION = PHRVA,D14,3251;%%
\bibitem [{\citenamefont {Sanchez}(1978)}]{Sanchez:1977si}%
  \BibitemOpen
  \bibfield  {author} {\bibinfo {author} {\bibfnamefont {N.G.} \bibnamefont
  {Sanchez}}, }\href {\doibase 10.1103/PhysRevD.18.1030} {\bibfield  {journal}
  {\bibinfo  {journal} {\emph {Phys. Rev.}} }\textbf {\bibinfo {volume} {D18}},
  \bibinfo {pages} {1030} (\bibinfo {year} {1978})}\BibitemShut {NoStop}%
%%CITATION = PHRVA,D18,1030;%%
\bibitem [{\citenamefont {Chandrasekhar}(1983)}]{MTB}%
  \BibitemOpen
  \bibfield  {author} {\bibinfo {author} {\bibfnamefont {S.}~\bibnamefont
  {Chandrasekhar}}, }\href@noop {} {\emph {\bibinfo {title} {The Mathematical
  Theory of Black Holes}}} (\bibinfo  {publisher} {Oxford University Press},
  \bibinfo {address} {New York}, \bibinfo {year} {1983})\BibitemShut {NoStop}%
\bibitem [{\citenamefont {Glampedakis} and \citenamefont
  {Andersson}(2001)}]{Glampedakis:2001cx}%
  \BibitemOpen
  \bibfield  {author} {\bibinfo {author} {\bibfnamefont {K.}~\bibnamefont
  {Glampedakis}} and \bibinfo {author} {\bibfnamefont {N.}~\bibnamefont
  {Andersson}}, }\href {\doibase 10.1088/0264-9381/18/10/309} {\bibfield
  {journal} {\bibinfo  {journal} {\emph {Class. Quant. Grav.}} }\textbf
  {\bibinfo {volume} {18}}, \bibinfo {pages} {1939} (\bibinfo {year} {2001})},
  \Eprint {http://arxiv.org/abs/gr-qc/0102100}
  {arXiv:gr-qc/0102100}\BibitemShut {NoStop}%
%%CITATION = GR-QC/0102100;%%
\bibitem [{\citenamefont {Macedo} \emph {et~al.}(2013)\citenamefont {Macedo},
  \citenamefont {Leite}, \citenamefont {Oliveira}, \citenamefont {Dolan}, and
  \citenamefont {Crispino}}]{Macedo:2013afa}%
  \BibitemOpen
  \bibfield  {author} {\bibinfo {author} {\bibfnamefont {C.F.B.} \bibnamefont
  {Macedo}}, \bibinfo {author} {\bibfnamefont {L.C.S.} \bibnamefont {Leite}},
  \bibinfo {author} {\bibfnamefont {E.S.} \bibnamefont {Oliveira}}, \bibinfo
  {author} {\bibfnamefont {S.R.} \bibnamefont {Dolan}},  and \bibinfo {author}
  {\bibfnamefont {L.C.B.} \bibnamefont {Crispino}}, }\href {\doibase
  10.1103/PhysRevD.88.064033} {\bibfield  {journal} {\bibinfo  {journal} {\emph
  {Phys. Rev.}} }\textbf {\bibinfo {volume} {D88}}, \bibinfo {pages} {064033}
  (\bibinfo {year} {2013})}, \Eprint {http://arxiv.org/abs/1308.0018}
  {arXiv:1308.0018}\BibitemShut {NoStop}%
%%CITATION = ARXIV:1308.0018;%%
\bibitem [{\citenamefont {Crispino} \emph {et~al.}(2009)\citenamefont
  {Crispino}, \citenamefont {Dolan}, and \citenamefont
  {Oliveira}}]{Crispino:2009xt}%
  \BibitemOpen
  \bibfield  {author} {\bibinfo {author} {\bibfnamefont {L.C.B.} \bibnamefont
  {Crispino}}, \bibinfo {author} {\bibfnamefont {S.R.} \bibnamefont {Dolan}},
  and \bibinfo {author} {\bibfnamefont {E.S.} \bibnamefont {Oliveira}}, }\href
  {\doibase 10.1103/PhysRevLett.102.231103} {\bibfield  {journal} {\bibinfo
  {journal} {\emph {Phys. Rev. Lett.}} }\textbf {\bibinfo {volume} {102}},
  \bibinfo {pages} {231103} (\bibinfo {year} {2009})}, \Eprint
  {http://arxiv.org/abs/0905.3339} {arXiv:0905.3339}\BibitemShut {NoStop}%
%%CITATION = ARXIV:0905.3339;%%
\bibitem [{\citenamefont {Leite} \emph {et~al.}(2016)\citenamefont {Leite},
  \citenamefont {Crispino}, \citenamefont {De~Oliveira}, \citenamefont
  {Macedo}, and \citenamefont {Dolan}}]{Leite:2016hws}%
  \BibitemOpen
  \bibfield  {author} {\bibinfo {author} {\bibfnamefont {L.C.S.} \bibnamefont
  {Leite}}, \bibinfo {author} {\bibfnamefont {L.C.B.} \bibnamefont {Crispino}},
  \bibinfo {author} {\bibfnamefont {E.S.} \bibnamefont {De~Oliveira}}, \bibinfo
  {author} {\bibfnamefont {C.F.B.} \bibnamefont {Macedo}},  and \bibinfo
  {author} {\bibfnamefont {S.R.} \bibnamefont {Dolan}}, }\bibfield  {booktitle}
  {\emph {\bibinfo {booktitle} {{Proceedings, 3rd Amazonian Symposium on
  Physics: Belem, Brazil, September 28-October 2, 2015}}}, }\href {\doibase
  10.1142/S0218271816410248} {\bibfield  {journal} {\bibinfo  {journal} {\emph
  {Int. J. Mod. Phys.}} }\textbf {\bibinfo {volume} {D25}}, \bibinfo {pages}
  {1641024} (\bibinfo {year} {2016})}\BibitemShut {NoStop}%
%%CITATION = IMPAE,D25,1641024;%%
\bibitem [{\citenamefont {Leite} \emph {et~al.}(2017)\citenamefont {Leite},
  \citenamefont {Dolan}, and \citenamefont {Crispino}}]{Leite:2017zyb}%
  \BibitemOpen
  \bibfield  {author} {\bibinfo {author} {\bibfnamefont {L.C.S.} \bibnamefont
  {Leite}}, \bibinfo {author} {\bibfnamefont {S.R.} \bibnamefont {Dolan}},  and
  \bibinfo {author} {\bibfnamefont {L.C.B.} \bibnamefont {Crispino}}, }\href
  {\doibase 10.1016/j.physletb.2017.09.048} {\bibfield  {journal} {\bibinfo
  {journal} {\emph {Phys. Lett.}} }\textbf {\bibinfo {volume} {B774}}, \bibinfo
  {pages} {130} (\bibinfo {year} {2017})}, \Eprint
  {http://arxiv.org/abs/1707.01144} {arXiv:1707.01144}\BibitemShut {NoStop}%
%%CITATION = ARXIV:1707.01144;%%
\bibitem [{\citenamefont {Leite} \emph {et~al.}(2018)\citenamefont {Leite},
  \citenamefont {Dolan}, and \citenamefont {Crispino}}]{Leite:2018mon}%
  \BibitemOpen
  \bibfield  {author} {\bibinfo {author} {\bibfnamefont {L.C.S.} \bibnamefont
  {Leite}}, \bibinfo {author} {\bibfnamefont {S.}~\bibnamefont {Dolan}},  and
  \bibinfo {author} {\bibfnamefont {C.B.} \bibnamefont {Crispino},
  \bibfnamefont {Luís}}, }\href {\doibase 10.1103/PhysRevD.98.024046}
  {\bibfield  {journal} {\bibinfo  {journal} {\emph {Phys. Rev.}} }\textbf
  {\bibinfo {volume} {D98}}, \bibinfo {pages} {024046} (\bibinfo {year}
  {2018})}, \Eprint {http://arxiv.org/abs/1805.07840}
  {arXiv:1805.07840}\BibitemShut {NoStop}%
%%CITATION = ARXIV:1805.07840;%%
\bibitem [{\citenamefont {Benone} \emph {et~al.}(2018)\citenamefont {Benone},
  \citenamefont {Leite}, \citenamefont {Crispino}, and \citenamefont
  {Dolan}}]{Benone:2018rtj}%
  \BibitemOpen
  \bibfield  {author} {\bibinfo {author} {\bibfnamefont {C.L.} \bibnamefont
  {Benone}}, \bibinfo {author} {\bibfnamefont {L.C.S.} \bibnamefont {Leite}},
  \bibinfo {author} {\bibfnamefont {C.B.} \bibnamefont {Crispino},
  \bibfnamefont {LuÍs}},  and \bibinfo {author} {\bibfnamefont {S.R.}
  \bibnamefont {Dolan}}, }\bibfield  {booktitle} {\emph {\bibinfo {booktitle}
  {{Proceedings, 4th Amazonian Symposium on Physics: Celebrating 100 years of
  the de Sitter solution and 60 years of Atsushi Higuchi: Belem, Brazil,
  September 18-22, 2017}}}, }\href {\doibase 10.1142/S0218271818430125}
  {\bibfield  {journal} {\bibinfo  {journal} {\emph {Int. J. Mod. Phys.}}
  }\textbf {\bibinfo {volume} {D27}}, \bibinfo {pages} {1843012} (\bibinfo
  {year} {2018})}, \Eprint {http://arxiv.org/abs/1809.08275}
  {arXiv:1809.08275}\BibitemShut {NoStop}%
%%CITATION = ARXIV:1809.08275;%%
\bibitem [{\citenamefont {Das} \emph {et~al.}(1997)\citenamefont {Das},
  \citenamefont {Gibbons}, and \citenamefont {Mathur}}]{Das:1996we}%
  \BibitemOpen
  \bibfield  {author} {\bibinfo {author} {\bibfnamefont {S.R.} \bibnamefont
  {Das}}, \bibinfo {author} {\bibfnamefont {G.W.} \bibnamefont {Gibbons}},  and
  \bibinfo {author} {\bibfnamefont {S.D.} \bibnamefont {Mathur}}, }\href
  {\doibase 10.1103/PhysRevLett.78.417} {\bibfield  {journal} {\bibinfo
  {journal} {\emph {Phys. Rev. Lett.}} }\textbf {\bibinfo {volume} {78}},
  \bibinfo {pages} {417} (\bibinfo {year} {1997})}, \Eprint
  {http://arxiv.org/abs/hep-th/9609052} {arXiv:hep-th/9609052}\BibitemShut
  {NoStop}%
%%CITATION = HEP-TH/9609052;%%
\bibitem [{\citenamefont {Zel'dovich}(1971)}]{zeldovich1}%
  \BibitemOpen
  \bibfield  {author} {\bibinfo {author} {\bibfnamefont {Y.B.} \bibnamefont
  {Zel'dovich}}, }\href@noop {} {\bibfield  {journal} {\bibinfo  {journal}
  {\emph {Pis'ma Zh. Eksp. Teor. Fiz.}} }\textbf {\bibinfo {volume} {14}},
  \bibinfo {pages} {270 [JETP Lett. {\bf14}, 180 (1971)]} (\bibinfo {year}
  {1971})}\BibitemShut {NoStop}%
\bibitem [{\citenamefont {Zel'dovich}(1972)}]{zeldovich2}%
  \BibitemOpen
  \bibfield  {author} {\bibinfo {author} {\bibfnamefont {Y.B.} \bibnamefont
  {Zel'dovich}}, }\href@noop {} {\bibfield  {journal} {\bibinfo  {journal}
  {\emph {Zh. Eksp. Teor. Fiz}} }\textbf {\bibinfo {volume} {62}}, \bibinfo
  {pages} {2076 [Sov.Phys. JETP {\bf 35}, 1085 (1972)]} (\bibinfo {year}
  {1972})}\BibitemShut {NoStop}%
\bibitem [{\citenamefont {Brito} \emph {et~al.}(2015)\citenamefont {Brito},
  \citenamefont {Cardoso}, and \citenamefont {Pani}}]{Brito:2015oca}%
  \BibitemOpen
  \bibfield  {author} {\bibinfo {author} {\bibfnamefont {R.}~\bibnamefont
  {Brito}}, \bibinfo {author} {\bibfnamefont {V.}~\bibnamefont {Cardoso}},  and
  \bibinfo {author} {\bibfnamefont {P.}~\bibnamefont {Pani}}, }\href {\doibase
  10.1007/978-3-319-19000-6} {\bibfield  {journal} {\bibinfo  {journal} {\emph
  {Lect. Notes Phys.}} }\textbf {\bibinfo {volume} {906}}, \bibinfo {pages}
  {pp.1} (\bibinfo {year} {2015})}, \Eprint {http://arxiv.org/abs/1501.06570}
  {arXiv:1501.06570}\BibitemShut {NoStop}%
%%CITATION = ARXIV:1501.06570;%%
\bibitem [{\citenamefont {Benone} and \citenamefont
  {Crispino}(2019)}]{Benone:2019all}%
  \BibitemOpen
  \bibfield  {author} {\bibinfo {author} {\bibfnamefont {C.L.} \bibnamefont
  {Benone}} and \bibinfo {author} {\bibfnamefont {L.C.B.} \bibnamefont
  {Crispino}}, }\href {\doibase 10.1103/PhysRevD.99.044009} {\bibfield
  {journal} {\bibinfo  {journal} {\emph {Phys. Rev.}} }\textbf {\bibinfo
  {volume} {D99}}, \bibinfo {pages} {044009} (\bibinfo {year} {2019})}, \Eprint
  {http://arxiv.org/abs/1901.05592} {arXiv:1901.05592}\BibitemShut {NoStop}%
%%CITATION = ARXIV:1901.05592;%%
\bibitem [{\citenamefont {Arvanitaki} \emph {et~al.}(2017)\citenamefont
  {Arvanitaki}, \citenamefont {Baryakhtar}, \citenamefont {Dimopoulos},
  \citenamefont {Dubovsky}, and \citenamefont {Lasenby}}]{Arvanitaki:2016qwi}%
  \BibitemOpen
  \bibfield  {author} {\bibinfo {author} {\bibfnamefont {A.}~\bibnamefont
  {Arvanitaki}}, \bibinfo {author} {\bibfnamefont {M.}~\bibnamefont
  {Baryakhtar}}, \bibinfo {author} {\bibfnamefont {S.}~\bibnamefont
  {Dimopoulos}}, \bibinfo {author} {\bibfnamefont {S.}~\bibnamefont
  {Dubovsky}},  and \bibinfo {author} {\bibfnamefont {R.}~\bibnamefont
  {Lasenby}}, }\href {\doibase 10.1103/PhysRevD.95.043001} {\bibfield
  {journal} {\bibinfo  {journal} {\emph {Phys. Rev.}} }\textbf {\bibinfo
  {volume} {D95}}, \bibinfo {pages} {043001} (\bibinfo {year} {2017})}, \Eprint
  {http://arxiv.org/abs/1604.03958} {arXiv:1604.03958}\BibitemShut {NoStop}%
%%CITATION = ARXIV:1604.03958;%%
\bibitem [{\citenamefont {Brito} \emph {et~al.}(2017)\citenamefont {Brito},
  \citenamefont {Ghosh}, \citenamefont {Barausse}, \citenamefont {Berti},
  \citenamefont {Cardoso}, \citenamefont {Dvorkin}, \citenamefont {Klein}, and
  \citenamefont {Pani}}]{Brito:2017wnc}%
  \BibitemOpen
  \bibfield  {author} {\bibinfo {author} {\bibfnamefont {R.}~\bibnamefont
  {Brito}}, \bibinfo {author} {\bibfnamefont {S.}~\bibnamefont {Ghosh}},
  \bibinfo {author} {\bibfnamefont {E.}~\bibnamefont {Barausse}}, \bibinfo
  {author} {\bibfnamefont {E.}~\bibnamefont {Berti}}, \bibinfo {author}
  {\bibfnamefont {V.}~\bibnamefont {Cardoso}}, \bibinfo {author} {\bibfnamefont
  {I.}~\bibnamefont {Dvorkin}}, \bibinfo {author} {\bibfnamefont
  {A.}~\bibnamefont {Klein}},  and \bibinfo {author} {\bibfnamefont
  {P.}~\bibnamefont {Pani}}, }\href {\doibase 10.1103/PhysRevLett.119.131101}
  {\bibfield  {journal} {\bibinfo  {journal} {\emph {Phys. Rev. Lett.}}
  }\textbf {\bibinfo {volume} {119}}, \bibinfo {pages} {131101} (\bibinfo
  {year} {2017})}, \Eprint {http://arxiv.org/abs/1706.05097}
  {arXiv:1706.05097}\BibitemShut {NoStop}%
%%CITATION = ARXIV:1706.05097;%%
\bibitem [{\citenamefont {Ikeda} \emph {et~al.}(2019)\citenamefont {Ikeda},
  \citenamefont {Brito}, and \citenamefont {Cardoso}}]{Ikeda:2019fvj}%
  \BibitemOpen
  \bibfield  {author} {\bibinfo {author} {\bibfnamefont {T.}~\bibnamefont
  {Ikeda}}, \bibinfo {author} {\bibfnamefont {R.}~\bibnamefont {Brito}},  and
  \bibinfo {author} {\bibfnamefont {V.}~\bibnamefont {Cardoso}}, }\href
  {\doibase 10.1103/PhysRevLett.122.081101} {\bibfield  {journal} {\bibinfo
  {journal} {\emph {Phys. Rev. Lett.}} }\textbf {\bibinfo {volume} {122}},
  \bibinfo {pages} {081101} (\bibinfo {year} {2019})}, \Eprint
  {http://arxiv.org/abs/1811.04950} {arXiv:1811.04950}\BibitemShut {NoStop}%
%%CITATION = ARXIV:1811.04950;%%
\bibitem [{\citenamefont {Abbott} \emph
  {et~al.}(2018)}]{LIGOScientific:2018mvr}%
  \BibitemOpen
  \bibfield  {author} {\bibinfo {author} {\bibfnamefont {B.P.} \bibnamefont
  {Abbott}} \emph {et~al.} (\bibinfo {collaboration} {LIGO Scientific, Virgo}),
  }\href@noop {} {  (\bibinfo {year} {2018})}, \Eprint
  {http://arxiv.org/abs/1811.12907} {arXiv:1811.12907}\BibitemShut {NoStop}%
%%CITATION = ARXIV:1811.12907;%%
\bibitem [{\citenamefont {Bernard} \emph {et~al.}(2019)\citenamefont {Bernard},
  \citenamefont {Cardoso}, \citenamefont {Ikeda}, and \citenamefont
  {Zilhão}}]{Bernard:2019nkv}%
  \BibitemOpen
  \bibfield  {author} {\bibinfo {author} {\bibfnamefont {L.}~\bibnamefont
  {Bernard}}, \bibinfo {author} {\bibfnamefont {V.}~\bibnamefont {Cardoso}},
  \bibinfo {author} {\bibfnamefont {T.}~\bibnamefont {Ikeda}},  and \bibinfo
  {author} {\bibfnamefont {M.}~\bibnamefont {Zilhão}}, }\href@noop {} {
  (\bibinfo {year} {2019})}, \Eprint {http://arxiv.org/abs/1905.05204}
  {arXiv:1905.05204}\BibitemShut {NoStop}%
%%CITATION = ARXIV:1905.05204;%%
\bibitem [{\citenamefont {Wong}(2019)}]{Wong:2019kru}%
  \BibitemOpen
  \bibfield  {author} {\bibinfo {author} {\bibfnamefont {L.K.} \bibnamefont
  {Wong}}, }\href@noop {} {  (\bibinfo {year} {2019})}, \Eprint
  {http://arxiv.org/abs/1905.08543} {arXiv:1905.08543}\BibitemShut {NoStop}%
%%CITATION = ARXIV:1905.08543;%%
\bibitem [{\citenamefont {Sommerfeld}(1964)}]{Sommerfeld:1964}%
  \BibitemOpen
  \bibfield  {author} {\bibinfo {author} {\bibfnamefont {A.}~\bibnamefont
  {Sommerfeld}}, }\href@noop {} {\emph {\bibinfo {title} {Optics; Lectures on
  Theoretical Physics, IV}}} (\bibinfo  {publisher} {Academic Press}, \bibinfo
  {address} {New York}, \bibinfo {year} {1964})\BibitemShut {NoStop}%
\bibitem [{\citenamefont {Restrick}(1968)}]{Restrick:1968}%
  \BibitemOpen
  \bibfield  {author} {\bibinfo {author} {\bibfnamefont {R.C.} \bibnamefont
  {Restrick}}, }\href@noop {} {\bibfield  {journal} {\bibinfo  {journal} {\emph
  {Radio Science}} }\textbf {\bibinfo {volume} {3}}, \bibinfo {pages} {1144}
  (\bibinfo {year} {1968})}\BibitemShut {NoStop}%
\bibitem [{\citenamefont {Merzbacher}(1998)}]{Merzbacher}%
  \BibitemOpen
  \bibfield  {author} {\bibinfo {author} {\bibfnamefont {E.}~\bibnamefont
  {Merzbacher}}, }\href@noop {} {\emph {\bibinfo {title} {Quantum Mechanics}}}
  (\bibinfo  {publisher} {John Wiley and Sons}, \bibinfo {address} {New York},
  \bibinfo {year} {1998})\BibitemShut {NoStop}%
\bibitem [{\citenamefont {Landau} and \citenamefont
  {Lifshitz}(1981)}]{landau1981quantum}%
  \BibitemOpen
  \bibfield  {author} {\bibinfo {author} {\bibfnamefont {L.}~\bibnamefont
  {Landau}} and \bibinfo {author} {\bibfnamefont {E.}~\bibnamefont {Lifshitz}},
  }\href@noop {} {\emph {\bibinfo {title} {Quantum Mechanics: Non-Relativistic
  Theory}}}, Course of Theoretical Physics (\bibinfo  {publisher} {Elsevier
  Science}, \bibinfo {year} {1981})\BibitemShut {NoStop}%
\bibitem [{\citenamefont {Zwillinger} \emph {et~al.}(2014)\citenamefont
  {Zwillinger}, \citenamefont {Moll}, \citenamefont {Gradshteyn}, and
  \citenamefont {Ryzhik}}]{tableofintegrals}%
  \BibitemOpen
  \bibinfo {editor} {\bibfnamefont {D.}~\bibnamefont {Zwillinger}}, \bibinfo
  {editor} {\bibfnamefont {V.}~\bibnamefont {Moll}}, \bibinfo {editor}
  {\bibfnamefont {I.}~\bibnamefont {Gradshteyn}},  and \bibinfo {editor}
  {\bibfnamefont {I.}~\bibnamefont {Ryzhik}}, eds., \href {\doibase
  https://doi.org/10.1016/B978-0-12-384933-5.00007-2} {\emph {\bibinfo {title}
  {Table of Integrals, Series, and Products}}}, \bibinfo {edition} {eighth} ed.
  (\bibinfo  {publisher} {Academic Press}, \bibinfo {address} {Boston},
  \bibinfo {year} {2014}) pp. \bibinfo {pages} {776 -- 865}\BibitemShut
  {NoStop}%
\bibitem [{\citenamefont {Campanelli} \emph
  {et~al.}(2007{\natexlab{a}})\citenamefont {Campanelli}, \citenamefont
  {Lousto}, \citenamefont {Zlochower}, and \citenamefont
  {Merritt}}]{Campanelli:2007ew}%
  \BibitemOpen
  \bibfield  {author} {\bibinfo {author} {\bibfnamefont {M.}~\bibnamefont
  {Campanelli}}, \bibinfo {author} {\bibfnamefont {C.O.} \bibnamefont
  {Lousto}}, \bibinfo {author} {\bibfnamefont {Y.}~\bibnamefont {Zlochower}},
  and \bibinfo {author} {\bibfnamefont {D.}~\bibnamefont {Merritt}}, }\href
  {\doibase 10.1086/516712} {\bibfield  {journal} {\bibinfo  {journal} {\emph
  {Astrophys. J.}} }\textbf {\bibinfo {volume} {659}}, \bibinfo {pages} {L5}
  (\bibinfo {year} {2007}{\natexlab{a}})}, \Eprint
  {http://arxiv.org/abs/gr-qc/0701164} {arXiv:gr-qc/0701164}\BibitemShut
  {NoStop}%
%%CITATION = GR-QC/0701164;%%
\bibitem [{\citenamefont {Bruegmann} \emph {et~al.}(2008)\citenamefont
  {Bruegmann}, \citenamefont {Gonzalez}, \citenamefont {Hannam}, \citenamefont
  {Husa}, and \citenamefont {Sperhake}}]{Brugmann:2007zj}%
  \BibitemOpen
  \bibfield  {author} {\bibinfo {author} {\bibfnamefont {B.}~\bibnamefont
  {Bruegmann}}, \bibinfo {author} {\bibfnamefont {J.A.} \bibnamefont
  {Gonzalez}}, \bibinfo {author} {\bibfnamefont {M.}~\bibnamefont {Hannam}},
  \bibinfo {author} {\bibfnamefont {S.}~\bibnamefont {Husa}},  and \bibinfo
  {author} {\bibfnamefont {U.}~\bibnamefont {Sperhake}}, }\href {\doibase
  10.1103/PhysRevD.77.124047} {\bibfield  {journal} {\bibinfo  {journal} {\emph
  {Phys. Rev.}} }\textbf {\bibinfo {volume} {D77}}, \bibinfo {pages} {124047}
  (\bibinfo {year} {2008})}, \Eprint {http://arxiv.org/abs/0707.0135}
  {arXiv:0707.0135}\BibitemShut {NoStop}%
%%CITATION = ARXIV:0707.0135;%%
\bibitem [{\citenamefont {Gonzalez} \emph {et~al.}(2007)\citenamefont
  {Gonzalez}, \citenamefont {Hannam}, \citenamefont {Sperhake}, \citenamefont
  {Bruegmann}, and \citenamefont {Husa}}]{Gonzalez:2007hi}%
  \BibitemOpen
  \bibfield  {author} {\bibinfo {author} {\bibfnamefont {J.A.} \bibnamefont
  {Gonzalez}}, \bibinfo {author} {\bibfnamefont {M.D.} \bibnamefont {Hannam}},
  \bibinfo {author} {\bibfnamefont {U.}~\bibnamefont {Sperhake}}, \bibinfo
  {author} {\bibfnamefont {B.}~\bibnamefont {Bruegmann}},  and \bibinfo
  {author} {\bibfnamefont {S.}~\bibnamefont {Husa}}, }\href {\doibase
  10.1103/PhysRevLett.98.231101} {\bibfield  {journal} {\bibinfo  {journal}
  {\emph {Phys. Rev. Lett.}} }\textbf {\bibinfo {volume} {98}}, \bibinfo
  {pages} {231101} (\bibinfo {year} {2007})}, \Eprint
  {http://arxiv.org/abs/gr-qc/0702052} {arXiv:gr-qc/0702052}\BibitemShut
  {NoStop}%
%%CITATION = GR-QC/0702052;%%
\bibitem [{\citenamefont {Campanelli} \emph
  {et~al.}(2007{\natexlab{b}})\citenamefont {Campanelli}, \citenamefont
  {Lousto}, \citenamefont {Zlochower}, and \citenamefont
  {Merritt}}]{Campanelli:2007cga}%
  \BibitemOpen
  \bibfield  {author} {\bibinfo {author} {\bibfnamefont {M.}~\bibnamefont
  {Campanelli}}, \bibinfo {author} {\bibfnamefont {C.O.} \bibnamefont
  {Lousto}}, \bibinfo {author} {\bibfnamefont {Y.}~\bibnamefont {Zlochower}},
  and \bibinfo {author} {\bibfnamefont {D.}~\bibnamefont {Merritt}}, }\href
  {\doibase 10.1103/PhysRevLett.98.231102} {\bibfield  {journal} {\bibinfo
  {journal} {\emph {Phys. Rev. Lett.}} }\textbf {\bibinfo {volume} {98}},
  \bibinfo {pages} {231102} (\bibinfo {year} {2007}{\natexlab{b}})}, \Eprint
  {http://arxiv.org/abs/gr-qc/0702133} {arXiv:gr-qc/0702133}\BibitemShut
  {NoStop}%
%%CITATION = GR-QC/0702133;%%
\bibitem [{\citenamefont {Sperhake} \emph {et~al.}(2011)\citenamefont
  {Sperhake}, \citenamefont {Berti}, \citenamefont {Cardoso}, \citenamefont
  {Pretorius}, and \citenamefont {Yunes}}]{Sperhake:2010uv}%
  \BibitemOpen
  \bibfield  {author} {\bibinfo {author} {\bibfnamefont {U.}~\bibnamefont
  {Sperhake}}, \bibinfo {author} {\bibfnamefont {E.}~\bibnamefont {Berti}},
  \bibinfo {author} {\bibfnamefont {V.}~\bibnamefont {Cardoso}}, \bibinfo
  {author} {\bibfnamefont {F.}~\bibnamefont {Pretorius}},  and \bibinfo
  {author} {\bibfnamefont {N.}~\bibnamefont {Yunes}}, }\href {\doibase
  10.1103/PhysRevD.83.024037} {\bibfield  {journal} {\bibinfo  {journal} {\emph
  {Phys. Rev.}} }\textbf {\bibinfo {volume} {D83}}, \bibinfo {pages} {024037}
  (\bibinfo {year} {2011})}, \Eprint {http://arxiv.org/abs/1011.3281}
  {arXiv:1011.3281}\BibitemShut {NoStop}%
%%CITATION = ARXIV:1011.3281;%%
\bibitem [{Dol()}]{Dolan:private}%
  \BibitemOpen
  \href@noop {} {}\bibinfo {note} {We thank Sam Dolan for pointing this
  parallel to us.}\BibitemShut {Stop}%
\bibitem [{Bec()}]{Beckmann}%
  \BibitemOpen
  \href@noop {} {}\bibinfo {note}
  {\url{https://eud.gsfc.nasa.gov/Volker.Beckmann/school/download/Longair_Radiation3.pdf}}\BibitemShut
  {NoStop}%
\end{thebibliography}%

\end{document}